\documentclass[conference]{IEEEtran}
\IEEEoverridecommandlockouts
\usepackage{cite}
\usepackage{amsmath,amssymb,amsfonts}
\usepackage{algorithmic}
\usepackage{graphicx}
\usepackage{subfigure,}
\usepackage{textcomp}
\usepackage{xcolor}
\usepackage{url}
\def\BibTeX{{\rm B\kern-.05em{\sc i\kern-.025em b}\kern-.08em
    T\kern-.1667em\lower.7ex\hbox{E}\kern-.125emX}}
\begin{document}

\title{Spreading processes with population heterogeneity over multi-layer networks\\
\thanks{This research was supported in part by the National Science Foundation through grants CCF-2225513 and CCF-1813637 and by the Army Research Office through grant \#W911NF-22-1-0181.     
}
}

\author{\IEEEauthorblockN{Yurun Tian}
\IEEEauthorblockA{\textit{Electrical and Computer Engineering} \\
\textit{Carnegie Mellon University}\\
Pittsburgh, USA \\
yurunt@andrew.cmu.edu}
\and
\IEEEauthorblockN{Osman Ya\u{g}an}
\IEEEauthorblockA{\textit{Electrical and Computer Engineering} \\
\textit{Carnegie Mellon Univeristy}\\
Pittsburgh, USA \\
oyagan@andrew.cmu.edu}
}

\maketitle

\begin{abstract}

It's been controversial whether re-opening school will facilitate viral spread among household communities with mitigation strategies such as mask-wearing in place. In this work, we propose an epidemiological model that explores the viral transmission over the multi-layer contact network composed of the school layer and community layer with population heterogeneity on mask-wearing behavior. We derive analytical expressions for three key epidemiological quantities: the probability of emergence, the epidemic threshold, and the expected epidemic size. In particular, we show how the aforementioned quantities depend on the structure of the multi-layer contact network, viral transmission dynamics, and the distribution of the different types of masks within the population. Through extensive simulations, our analytical results show near-perfect agreement with the simulation results with a limited number of nodes. Utilizing the model, we study the impact of the opening/closure of the school layer on the viral transmission dynamics with various mask-wearing scenarios. Interestingly, we found that it's safe to open the school layer with the proper proportion of good-quality masks in the population. Moreover, we validate the theory of the trade-off between source-control and self-protection over a single layer by Tian et al on our multi-layer setting. We conclude that even on a multi-layer network, it's of great significance to treat the spreading process as two distinct phases in mind when considering mitigation strategies. Besides, we would like to remark that our model of spreading process over multi-layer networks with population heterogeneity can also be applied to various other domains, such as misinformation control.

\end{abstract}


\section{Introduction}



School closures have been one of the main non-pharmaceutical interventions to suppress the the epidemics of droplet infections such as COVID-19 \cite{WHO_public_2021}, and have shown effectiveness in cases \cite{brooks_impact_2020}. 
It has been controversial that whether re-opening school for in-person instruction will facilitate disease transmission during the COVID-19 pandemic \cite{viner_school_2020}. 
There are studies showing that the school will bring more risks to the household transmission, due to the potentially infectious contacts increase through children mixing in school, as well as the fact that children link the contact networks of schools and households \cite{sadique_estimating_2008}.
On the other hand, some studies argue that with mask wearing and physical distancing strategies enforced, asymptomatic infection rates appeared low, indicating that it might be safe to open school with appropriate mitigation efforts in place \cite{katz_low_2021}.  
Despite many sources of reflection, a univocal and certain answer to this question has not clearly emerged \cite{nenna_weighing_2022}. 
Qualitative analysed studies deeply differ and are hardly comparable, regarding the size of involved subjects (varies from hundreds to thousands), the time length of the studies, and SARS-Cov-2 screening guidelines, etc \cite{busa_covid-19_2021}.

Epidemiological models that analyze the transmission dynamics of viral spread provide flexibility for stimulating and predicting under certain scientifically-controlled conditions \cite{iranzo_epidemiological_2021}. 
Within the context of school closure and reopening with mitigation strategies engaged (e.g., mask-wearing), 
some work considers scenarios where individuals wear different types of masks on a single-layer of contact networks \cite{eikenberry_mask_2020, sridhar_leveraging_2021, tian_analysis_2021, yagan_special_2020}.
However, single-layer \textit{mask model} is not able to model real-life situations where the virus transmits in between two distinct contact networks.
For example, students at school obtain infection while sitting together in the classroom, and transmit the infection to their family members when having dinner, or vice versa. 
In order to capture the distinct viral transmission dynamics on different layers of contact networks, it's necessary to consider multi-layer rather than a single-layer contact networks.
Other work using epidemiological models take into account multi-layer  networks but no population heterogeneity on mitigation effect considered \cite{bianconi_epidemic_2018, chen_multilayer_2022, salehi_spreading_2015, brodka_interacting_2020, yi_multilayer_2022}.
\cite{espana_impacts_2021} takes into account both multi-layer network and mask-wearing with a agent-based simulation-only framework that incorporates fine-grained real-life interaction details (e.g., household location) for K-12 school reopening in Indiana, but does not provide a analytical backbone that provides mathematical tractability and prone to the data collection quality \cite{shi_agent-based_2014}.

In order to explore the impact of opening and closure of schools to the viral transmission dynamics with mitigation strategies (e.g., mask-wearing) considered from a principled, mathematical lens, 
we propose a \textit{multi-layer mask model} for analyzing and predicting viral spreading dynamics over multi-layer networks.
Our model is based on the \textit{mask model} proposed in \cite{tian_role_2022},
where population heterogeneity of the individual-level mitigation effect of viral transmission ( e.g., personal mask choice) on a single contact network is taken into account. 
\textit{Mask model} assumes there are $M$ types of masks exists in the population.
Different masks exhibit various inward and outward filtration efficiencies, resulting in heterogeneous transmissibilities between infectious and susceptible node pairs. 
No-mask is included as one of the $M$ mask types only without any filtration efficiency. 
The population heterogeneity paradigm proposed by \textit{mask model} can also adapt to other application contexts such as misinformation spreading, where individuals exhibit various tendencies to accept and transmit information based on their personalities or fact-checking behaviors.
Interestingly, in \cite{tian_role_2022}, the results emphasize the trade-off between source-control and self-protection: spreading process should be treated as two distinct stages in mind, i.e., before the epidemic exists, mitigation strategies should focus on source-control; while  if epidemic already emerges, it's more effective to focus on self-protection to suppress further expansion of existing epidemic.

Our contributions are twofold.
First, we extend the \textit{mask model} from the single-layer setting to multi-layer, and derive \textit{analytical predictions} for three important epidemiological quantities: the probability of emergence (PE), the epidemic threshold and expected epidemic size (ES). Specifically, we show how these quantities depend on the structure of the multi-layer network, viral spread dynamics, as well as the mask distribution within the population. 
The results are built upon the theory of multi-type and multi-layer branching processes \cite{brodka_interacting_2020, athreya_t_2011, yagan_conjoining_2013}. 
Second, we validate our analytical results through extensive simulations and demonstrate that our predictions are in good agreement with empirical results. 
First, we study the spreading process as a function of mean degrees of community network and school network.
The results reveal the interplay between network structure and spreading process dynamics, and demonstrated the effectiveness of preventing the epidemic or suppressing the expansion if the epidemic already exists by reducing mean degrees 
Next, we explore the impact of open/closure of the school layer with various mask-wearing distribution. 
We found that with proper mask-wearing mitigation in place, it's safe to open the school layer without triggering the epidemic. 
Last but not the least, we conduct experiments to validate the result on trade-off between source-control and self-protection on the multi-layer network setting by \cite{tian_role_2022}.

To this end, we would like to remark that, though this paper is mainly focusing on the impact of open and closure of the school layer to the viral transmission,  joint consideration of population heterogeneity with multi-layer \textit{information contact networks} is of great significance under the context of information diffusion. The emergence of new form of augmented reality platforms such as Metaverse has drawn great attention on information spreading on multi-layer social networks \cite{noauthor_physical_nodate, moro-visconti_metaverse_2022}. 
To our best knowledge, there is little prior work on spreading processes on multi-layer networks with population heterogeneity. Notable ones are \cite{allard_heterogeneous_2009, lee_epidemic_2021, tian_role_2022, tian_analysis_2021}.

\section{Model}

\subsection{Multi-layer contact network model}
\label{sec:network_model}
We construct our school-community two-layer contact network for viral spread based on the multi-layer network model considered by Ya\u{g}an et al in \cite{yagan_conjoining_2013}. 
Consider a population of size $n$ with individuals in the set $\mathcal{N}=\{1, \ldots, n\}$. 
Within the multi-layer contact network, each node correspond to an individual in $\mathcal{N}$, and an edge is drawn between two nodes if they are come in contact by which the virus can potentially transmit.
Let $\mathbb{C}$ stand for the community contact network of individuals on the node set $\mathcal{N}$. 
Let $\mathbb{S}$ represent the school contact network with the assumption that each node in $\mathcal{N}$ is a member of the school network $\mathbb{S}$ with probability $\alpha \in (0, 1]$, independently from any other node. 
Formally, we let 

\begin{equation}
\mathbb{P}\left[i \in \mathcal{N}_S\right]=\alpha, \quad i=1, \ldots, n    
\end{equation}

where $\mathcal{N}_S$ denote the set of individuals who also attend the school layer. 
Edges belonging to network $\mathbb{C}$ (resp. $\mathbb{S}$) are noted as type-$c$ (resp. type-$s$) edges, representing viral transmission paths via community (reps. school) contacts. 

In line with prior literature on stochastic epidemic models, we generate the contact networks $\mathbb{C}$ and $\mathbb{S}$ by the \textit{configuration model} \cite{molloy_critical_nodate}. 
The topology of the networks $\mathbb{C}$ and $\mathbb{S}$ are decided via their respective degree distributions $\{p_k^c\}$ and $\{p_k^s\}$ where $k \geq 0$, respectively. 
In other words, $p_k^c$ (resp. $p_k^s$) is the probability that an arbitrary node on network $\mathbb{C}$ (resp. $\mathbb{S}$) has degree $k$, i.e., it is connected to $k$ other nodes via an \textit{undirected} type-$c$ (resp. type-$s$) edge. 
We generate networks $\mathbb{C} (n; {p_k^c})$ and $\mathbb{S} (n; 
\alpha, {p_k^s})$ according to the configuration model separately.
Based on the configuration model, the degrees of nodes in network $\mathbb{C}$ and $\mathbb{S}$ are drawn independently from their own degree distributions $\{p_k^c\}$ and $\{p_k^s\}$.
Here we shall assume that all moments of arbitrary order are finite for these degree distributions.

In order to model viral transmission among individuals through both layers, we characterize an overlay network $\mathbb{H}$ by taking the union of $\mathbb{C}$ and $\mathbb{S}$, i.e., $\mathbb{H} = \mathbb{C} \cup \mathbb{S} $. 
In other words, for any pair of nodes $i$, $j$, we say that $i$ and $j$ are adjacent in the network $\mathbb{H}$ as long as there is at least an edge connecting them via either an type-$c$ or type-$s$ edge. 
In this setting, the \textit{colored} degree of a node $i$ is represented by an integer vector $\boldsymbol{d^i}=\left[k_c^i, k_s^i\right]$, where $k_c^i$ (resp. $k_s^i$) stands for the number of community edges (resp. school edges). 
Given $\{p_k^c\}$ and $\{p_k^s\}$, and consider the independence of $\mathbb{C}$ and $\mathbb{S}$, we  formulate the colored degree distribution as follow $p_{\boldsymbol{d}}$ as:
\vspace{-1mm}
\begin{equation}
    p_{\boldsymbol{d}}=\left(\alpha p_{k_c}^c+(1-\alpha) \mathbf{1}\left[k_s=0\right]\right) \cdot p_{k_s}^s, \quad \boldsymbol{d}=\left(k_c, k_s\right)
\end{equation}

where, the term $(1-\alpha)\mathbf{1}\left[k_s=0\right]$ accounts for the case where the node is not a member of the school layer $\mathbb{S}$, and the number of type-$s$ edges is automatically zero.
If $\sum_{i=1}^n k_c^i$ and $\sum_{i=1}^n k_s^i$ are even, we construct $\mathbb{H}$ as in [conjoining 40 and 21]. 
Specifically, each node $i = 1, ..., n$ is first given the $k_c^i$ and $k_s^i$ stubs of type-$c$ and type-$s$, respectively. 
Then, the stub pairs of the same link type are randomly picked and connected together to form complete edges. The process of pairing stubs continues until none is left.



\subsection{Transmission model}
In the seminal work \cite{newman_spread_2002}, Newman studied an SIR (susceptible-infectious-recovered) model of viral spread over a contact network.
Simple and generalizable, Newman's model captures complex viral transmission and recovery mechanisms via proper selection of $T$ \cite{newman_spread_2002}.
Many authors have incorporated various agent-level heterogeneity based on Newman's model 
\cite{alexander_risk_2010, allard_heterogeneous_2009, eletreby_effects_2020, sridhar_leveraging_2021, tian_analysis_2021, yagan_conjoining_2013, yagan_special_2020, tian_role_2022}, 
among which \textit{mask model} \cite{tian_role_2022, tian_analysis_2021} presents a direct modeling of population heterogeneity via inward efficiency and outward efficiency of masks with accurate predictions on three epidemiological quantities: probability of emergence, epidemic threshold and expected epidemic size. 
Specifically, 
inward-efficiency is the probability that respiratory droplets will be blocked from the outside layer of the mask to the inside, which quantifies the protection of the mask against receiving the virus.
Outward efficiency is the probability that respiratory droplets will be kept from the inside layer of the mask to the outside, quantifying the protection against transmitting the virus. 
In \textit{mask model}, given the original transmissibilty of the virus $T$, the transmission probability from a type-$i$ individual to a type-$j$ individual is given by
\vspace{-1mm}
\begin{equation*}
    \mathbf{T}{[i, j]}:=\left(1-\epsilon_{ {out }, i}\right)\left(1-\epsilon_{{in }, j}\right) T, 
    \; 1 \leq i, j \leq M
\end{equation*}

where $\epsilon_{out, i}$ is the outward efficiency of a type-$i$ mask, $\epsilon_{in, j}$ is the inward-efficiency of a type-$j$ mask, $M$ is the total number of  mask types and $T$ as the original transimissibility of the virus without any masks.
We have $\epsilon_{\text {out }, i}, \epsilon_{i n, j} \in[0,1], \forall i, j \in[1, M]$, and $\mathbf{T}$ is a $M \times M$ transmissibility matrix. 
In this work, in line with \textit{mask model}, population heterogeneity is modeled by the $M$ types of masks that nodes wear.
In addition, mask disitrbution is described by $\{m_1, ..., m_M\}$ over the set ${1, ..., M}$ where $m_i$ represents the fraction of individuals who wear a mask of type-$i$.
Another assumption is the type of mask is pre-assigned independently from 
$\left\{m_i\right\}_{i=1}^M$ over all nodes in the network before the spreading process starts.
For notational convenience, we shall say that an individual is of type-$i$ if they wear a type-$i$ mask ($1 \leq i \leq M$). 

Including \textit{mask model}, these models only consider the single-layer setting for the heterogeneous bond percolation process, where the likelihood of infection depends only on the node types but \textit{not} the link type over which transmission occurs.
We extend the \textit{mask model} on a single-layer to multi-layer to account for the fact that the viral transmission is not only governed by one type of transmission pathways \cite{leung_transmissibility_2021}. 
As discussed in Section \ref{sec:network_model}, the multi-layer network $\mathbb{H}$ is composed of two network layers $\mathbb{C}$ and $\mathbb{S}$. 
In our \textit{multi-layer mask model}, we assume that the \textit{transmissibility}, i.e., the probability that an infected individual passes on the infection to their direct contacts depends on \textit{both} the type of masks they are wearing, and the type of links connecting them.
We assume the original transmissibility of the virus is $T_c$ on type-$c$ links and $T_s$ on type-$s$ links. 
Compared to \textit{mask model}, instead of one $M \times M$  transmissibility matrix for a single layer, now we have two transmissibility matrices $\mathbf{T_c}$ and $\mathbf{T_s}$ each of size $M \times M$ for layer $\mathbb{C}$ and $\mathbb{S}$ separately. 
More specifically, assume that each edge in $\mathbb{C}$ can transmit the pathogen with probability
\begin{equation*}
 \mathbf{T_c}{[i, j]} := T_c^{i, j} = \left(1-\epsilon_{ {out }, i}\right)\left(1-\epsilon_{{in }, j}\right) T_c, 
    \; 1 \leq i, j \leq M
\end{equation*}

each edge in $\mathbb{S}$ can transmit the pathogen with probability
\vspace{-1mm}
 \begin{equation*}
 \mathbf{T_s}{[i, j]} := T_s^{i, j} =\left(1-\epsilon_{ {out }, i}\right)\left(1-\epsilon_{{in }, j}\right) T_s, 
    \; 1 \leq i, j \leq M
\end{equation*}
 if the infected node at the end of the edge wears a type-$i$ mask and the susceptible node at the other end wears a type-$j$ mask.


\section{Analysis Results}

\subsection{Probability of Emergence and Epidemic Threshold}

Consider random graphs $\mathbb{C}(n, \{p_k^c\})$ and $\mathbb{S}(n; \alpha, \{p_k^s\})$ as in Section \ref{sec:network_model}. 
In order to study the viral transmission in the multi-layer network $\mathbb{H} = \mathbb{C} \cup \mathbb{S}$ where nodes wear different types of masks, we consider a branching process which starts by giving the pathogen to an arbitrary node, and next, we recursively discover the largest number of nodes that are reached and \textit{infected} by exploring its neighbors.
We consider the scenario where 
the virus transmits from a type-$i$ infected node to a type-$j$ susceptible neighbor node via a  type-$c$ link with probability ${T_c}^{i,j} = T_c (1 - \epsilon_{i, out}) (1 - \epsilon_{j, in})$ (or through a type-$s$ link with probability $ {T_s}^{i,j} = T_s (1 - \epsilon_{i, out}) (1 -  \epsilon_{j, in}$), independently from all the other neighbors.
Here we leverage the standard approach on generating functions \cite{newman_random_2001, newman_spread_2002} with consideration of node types to determine the condition for the existence of a giant infected component induced by an arbitrary initial infected node.
Note it's required that the initial stages of the branching processes is locally tree-like. This is promised as the clusering coefficient of colored degree-driven networks scales like $1/n$ as $n$ approximates infinity \cite{soderberg_properties_2003}.

The survival probability of the aforementioned branching process is solved via the \textit{mean-field} approach on the generating function \cite{newman_random_2001, newman_spread_2002}.
For $1 \leq i \leq M$, let $h_{c,i} (x)$ (resp. $h_{s, i} (x)$) denote the generating functions for the \textit{finite} number of nodes reached and infected by following an type-$c$ (resp. type-$s$) edge coming from a type-$i$ infected node. 
Put differently, in the probability generating function $h_{c, i} (x) = \sum_{n=0}^{\infty} v_n x^n$, $v_n$ denotes the probability that an arbitrary type-$c$ edge coming from a type-$i$ infected node infected a component of \textit{finite} size $n$. Similar to $h_{s, i} (x)$.
Besides, let $H_i(x)$ denotes the generating function for the \textit{finite} number of nodes infected from an arbitrary infected type-$i$ node.

Now we derive the recursive formulation of $h_{c,i} (x)$ as a function of $h_{c,1} (x), ..., h_{c,M} (x)$ and $h_{s,1} (x), ..., h_{s,M} (x)$. The derivation of $h_{s,i} (x)$ can be done analogously.
Given an infected type-$i$ node $v$ with an type-$c$ edge coming out of it, the first thing is to decide what type of node will be reached by the edge. 
Since the mask assignment is done before the spreading process and independently for all the nodes, by probability $m_j$ will we meet a type-$j$ susceptible neighbor, say node $u$. 
Next, the infected node $v$ tries to transmit the virus to susceptible neighbor $u$ with mask-reduced transmissibility $\mathbf{T_c}^{ij} = T_c (1 - \epsilon_{out, i}) (1 - \epsilon_{in, j})$. 
If the transmission fails, node $v$ will have zero offspring by reaching out to $u$. If the transmission succeeds, the number of infected neighbors induced by node $v$ will increase by one, and we need to continue to collect all the nodes infected by $u$. 
The factor ${p_{\boldsymbol{d}} k_c}/{\left\langle k_c\right\rangle}$ provides \cite{newman_spread_2002} the \textit{normalized} probability that an edge of type-$c$ is attached to a node at the other end with colored degree $\boldsymbol{d} = (k_c, k_s)$. 
Since node $u$ is reached by $v$ with a type-$c$ edge, it can infect other nodes via its remaining $k_c - 1$ links of type-$c$ and $k_s$ links of type-$c$.
Given that the number of nodes reached and infected by each of its type-$c$ (resp. type-$s$) links is generated by $h_{c, j} (x)$ (resp. $h_{s, j} (x)$), by the powers property of generating functions \cite{newman_random_2001}, we can obtain the term ${h_{c, j}(x)}^{k_c - 1} {h_{s, j}(x)}^{k_s}$ (resp. ${h_{s, j}(x)}^{k_s - 1} {h_{c, j}(x)}^{k_c}$) to describe the total number of nodes reached and infected by node $u$. 
Put together, we have
\begin{equation}
\small
\label{eq:hci}
    h_{c,i} = \sum_{j=1}^M m_j [1 - T_{c, ij} + T_{c, ij} x \sum_{\boldsymbol{d}} \frac{p_{\boldsymbol{d}} k_c}{\left\langle k_c\right\rangle} {h_{c, j}(x)}^{k_c - 1} {h_{s, j}(x)}^{k_s}]
\end{equation}
\vspace{-1mm}
\begin{equation}
\small
\label{eq:hsi}
    h_{s,i} = \sum_{j=1}^M m_j [1 - T_{s, ij} + T_{s, ij} x \sum_{\boldsymbol{d}} \frac{p_{\boldsymbol{d}} k_s}{\left\langle k_s\right\rangle} {h_{s, j}(x)}^{k_s - 1} {h_{c, j}(x)}^{k_c}]
\end{equation}

Utilizing equation (\ref{eq:hci}) and (\ref{eq:hsi}), we now derive the \textit{finite} number of nodes reached and infected by the aforementioned branching process. Similarly we have

\begin{equation}
\label{eq:H}
H_i(x) = x \sum_{\boldsymbol{d}} p_{\boldsymbol{d}} {h_{c, i}(x)}^{k_c} {h_{s, i}(x)}^{k_s}
\end{equation}

The factor $x$ corresponds to the initial node that is selected arbitrarily and infected. 
The selected node has colored degree $\boldsymbol{d} = (k_c, k_s)$ with probability $p_{\boldsymbol{d}}$. The number of nodes it reaches and infects by each of its $k_c$ (resp. $k_s$) links of type-$c$ (resp. type-$s$) is generated through $h_{c, i} (x)$ (resp. $h_{s, i} (x)$). Take average over all the possible colored degrees, we obtain equation (\ref{eq:H}).

With equation (\ref{eq:hci}) to (\ref{eq:H}) in hand, the desired generating function $H_i(x)$ for the finite number of nodes can be computed in the following manner. 
Given any $x$ value, we solve the recursive relations and will obtain the value for ${h_c^1} (x), ..., {h_c^M} (x)$, ${h_s^1} (x), ..., {h_s^M} (x)$ and $H_1 (x), ..., H_M (x)$. 
If we repeat the same process for any $x$ value, we will obtain a complete characterization of $H_1 (x), ..., H_M (x)$.
However, we are only interested in the cases where the number of nodes reached and infected by the initial node is \textit{infinite}, which represents the cases where a random chosen infected node triggers an \textit{epidemic}.
More precisely, we aim to derive the 
i) conditions in which there is a non-zero probability of infecting a positive fraction of nodes as n approximates infinity;
and ii) the exact value of this non-zero probability. 

The \textit{conservation of probability} property of generating functions indicates that there exists a trivial fixed point $h_{c,i} (1) = h_{s,i} (1) = 1$ (yielding $H_i(1) = 1$) when the number of nodes reached and infected is always \textit{finite}. In other words, the underlying branching process is in the sub-critical regime and \textit{all} infected components have finite sizes.
However, the fixed point $h_{c,i} (1) = h_{s,i} (1) = 1$ may not be a stable solution to the recursion (\ref{eq:hci}) to (\ref{eq:H}). 
We can check the stability of this fixed point by the
linearization of recursion (\ref{eq:hci}) to (\ref{eq:H}) around $h_{c,i} (1) = h_{s,i} (1) = 1$.
This yields the Jacobian matrix $\boldsymbol{J}$ with the form
\vspace{-1mm}
\begin{equation}
    \boldsymbol{J}=\left[\begin{array}{cc}
\boldsymbol{J}_{cc} & \boldsymbol{J}_{cs} \\
\boldsymbol{J}_{sc} & \boldsymbol{J}_{ss}\\

\end{array}\right]_{2M \times 2M}
\end{equation}

where

\begin{equation}
\tiny
    \begin{aligned}
    & \boldsymbol{J}_{cc}(i, j)=\left.\frac{\partial h_{c,i}(1)}{\partial h_{c,j}(1)}\right|_{h_{c,i}(1)=h_{c,j}(1)=1} 
    \boldsymbol{J}_{cs}(i, j)=\left.\frac{\partial h_{c,i}(1)}{\partial h_{s,j}(1)}\right|_{h_{c,i}(1)=h_{s,j}(1)=1}\\
    & \boldsymbol{J}_{ss}(i, j)=\left.\frac{\partial h_{s,i}(1)}{\partial h_{s,j}(1)}\right|_{h_{s,i}(1)=h_{s,j}(1)=1}
    \boldsymbol{J}_{sc}(i, j)=\left.\frac{\partial h_{s,i}(1)}{\partial h_{c,j}(1)}\right|_{h_{s,i}(1)=h_{c,j}(1)=1}
    \end{aligned}
\end{equation}
If all eigenvalues of $\boldsymbol{J}$ are less than one in absolute value, i.e., the spectral radius $\rho(\boldsymbol{J})$ of $\boldsymbol{J}$ satisfies $\rho(\boldsymbol{J}) \leq 1$, then the solution $h_{c,i} (1) = h_{s,i} (1) = 1$ is stable and $H_i(1) = 1$ becomes the physical solution.
This means that $\mathbb{H} = \mathbb{C} \cup \mathbb{S}$ does not possess a giant component with high probability.
In this case, as $n$ approximates infinity, the fraction of nodes that are infected tends to be zero. 
In contrast, if $\rho(\boldsymbol{J}) > 1$, the the fixed point is not stable, which indicates that the asymptotic branching process is supercritical. In this scenario, there is a positive probability of producing infinite trees.
A nontrivial fixed point exists and becomes the attractor of the recursions  (\ref{eq:hci}) to (\ref{eq:H}), leading to a solution with $h_{c,i} (1), h_{s,i} (1) < 1$ and further implies $H_i(1) < 1$. Thus the probability that there is a giant (\textit{infinite}) component of infected nodes can be attributed by $1 - H_i(1)$ for a type-$i$ initiator. 
The average probability of emergence with a random initiator is of form $\sum_{i=i}^M m_i (1 - H_i(1))$.
Collecting, the epidemic threshold is given by $\rho(\boldsymbol{J}) = 1$. To conserve space we do not provide the exact mathematical format of $\boldsymbol{J}$ which can be easily derived by elementary algebra or obtained numerically via off-the-shelf Jacobian matrix calculation tools such as the one provided by PyTorch \cite{torch_j}.

\subsection{Epidemic Size}
In this section, we compute the expected size of the epidemic, i.e., the final fraction of infected individuals of each type, conditioned on the event that the epidemic does not die out in finite time. We follows work by \cite{eletreby_effects_2020, gleeson_cascades_2008, gleeson_seed_2007, newman_epidemics_2018,tian_analysis_2021}.
Since the multi-layer network $\mathbb{H}$ is locally tree-like as the network size approximates infinity \cite{soderberg_properties_2003}, we consider a hierarchical structure, where at the top level, there is a single node (the \textit{root}). 
The probability that an arbitrarily selected root node is infected equals the value for fraction of individuals infected by the virus. We compute the probability that a randomly chosen root is infected by considering the tree-like neighborhood around it. 
We label the levels of the tree from the bottom level $L=0$ to the top level $L = \infty$ where the root is at. 
Let $q_{c(s), \ell}^i$ denotes the probability of a type-$i$ node at level $\ell$ is \textbf{\textit{not}} infected and connects to a node at level $\ell+1$ with a type-$c$ ($s$) edge.
Let $q_{\infty}^i$ be the probability of a type-$i$ node is \textit{not} infected, i.e., the expected fraction of individuals infected wearing type-$i$ masks. 
Note that no mask can also be regarded as a type of mask but with zero filtration efficiency. 
Next we will describe the process of computing $q_{c(s), \ell}^i$ in a recursive manner.

Let $u$ be a given type-$i$ node at level $\ell+1$ with an edge of type-$c$ connecting to a node at level $\ell + 2$, and $q_{c, \ell+1}^i$ is the probability that $u$ is \textit{not} infected and connects to a node at level $\ell+2$ via a type-$c$ link. 
Let $k_{c}^1, ..., k_{c}^M$ (resp. $k_{s}^1, ..., k_{s}^M$) and
 denote the number of neighbors from level $\ell$ of each type that connect to node $u$ via type-$c$ (resp. type-$s$) links.
 Let $X_{c}^1, ..., X_{c}^M$ (resp. $X_{s}^1, ..., X_{s}^M$) denote the number of \textit{infected} level $\ell$ neighbors of each type that connect to node $u$ via type-$c$ (resp. type-$s$) links. 
 Conditioned on $X_{c}^1, ..., X_{c}^M$ and $X_{s}^1, ..., X_{s}^M$, the probability that node $u$ is \textit{not} infected is $ \prod_{j = 1}^{M} {(1 - {T_c}^{j,i})}^{X_{c}^j}{(1 - {T_s}^{j,i})}^{X_{s}^j}$.

 Next, we take an expectation over the $X_c^j$ and $X_s^j$ to compute the unconditional probability of non-infection.
 Note that the infection status of nodes at the same level are independent from each other considering nodes in a given level do not share common infected ancestors due to the locally tree-like structure of the network. 
 Therefore we have 
 $X_c^j \sim \operatorname{Binomial}\left(k_c^j, 1 - q_{c,\ell}^j\right)$ and 
 $X_s^j \sim \operatorname{Binomial}\left(k_s^j, 1 - q_{s,\ell}^j\right)$, and $X_c^j$ and $X_s^j$ are independent from each other, too.

The probability of $u$ \textit{not} being infected condition on $k_{c}^1, ..., k_{c}^M$ and $k_{s}^1, ..., k_{s}^M$ is therefore given by
\begin{equation}
\small
\label{eq:condition_nodetype_linktype}
\begin{aligned}
& \mathbb{E}\left[\prod_{j = 1}^{M} {(1 - {T_c}^{j,i})}^{X_c^j}{(1 - {T_s}^{j,i})}^{X_s^j} \mid k_{c}^1, ..., k_{c}^M, k_{s}^1, ..., k_{s}^M \right] \\
&=\prod_{j=1}^M
\left(
1 - {T_c}^{j,i} + q_{c, \ell}^j {T_c}^{j,i}
\right)^{k_c^j} \left(1 - {T_s}^{j,i} + q_{s, \ell}^j {T_s}^{j,i})\right)^{k_s^j} \\
\end{aligned}
\end{equation}

Our next step is to condition on $u$'s total number of neighbors at level $\ell$ that reach $u$ via type-$c$ and type-$s$ links separately, denoted by $k_c = k_c^1 + ... + k_c^M$ and $k_s = k_s^1 + ... + k_s^M$.
The tuple $(k_c^1, ..., k_c^M)$ is drawn from the $\operatorname{Multinomial}\left(k_c, m_1, \ldots, m_M\right)$ distribution, 
and $(k_c^1, ..., k_c^M)$ is drawn from 
$\operatorname{Multinomial}\left(k_s, m_1, \ldots, m_M\right)$.
In other words, the probability of observing a instantiation 
$(k_c^1, ..., k_c^M)$ 
is $ \left(\begin{array}{c}
k_c \\
k_c^1, \ldots, k_c^M
\end{array}\right) m_1^{k_c^1} \ldots m_M^{k_c^M} $. $k_s$ shares the similar format but omitted to conserve space.


Next, we can rewrite equation (\ref{eq:condition_nodetype_linktype}) to be conditioned on $k_c$ and $k_s$ as:

\vspace{-1mm}
\begin{equation}
\small
    \begin{aligned}
& \left(\sum_{j=1}^M m_j (1 - T_c^{j,i} + q_{c, \ell}^j T_c^{j,i})
\right)^{k_c} \left(\sum_{j=1}^M m_j (1 - T_s^{j,i} + q_{s, \ell}^j T_s^{j,i})
\right)^{k_s}\\
& = f_i(\boldsymbol{q}_{c,\ell}, \boldsymbol{q}_{s,\ell}, k_c, k_s)
    \end{aligned}
\end{equation}

where we use $\boldsymbol{q}_{c,\ell} = [q_{c, \ell}^1, q_{c, \ell}^2, ..., q_{c, \ell}^M]$, 
$\boldsymbol{q}_{s,\ell} = [q_{s, \ell}^1, q_{s, \ell}^2, ..., q_{s, \ell}^M]$ and $f_i$
for simplicity of mathematical representation.
Finally, to remove the conditioning on the number of neighbors coming from each type of links, i.e., $k_c$ and $k_s$, we take expectation over the colored excess degree distribution of $u$. 
We use $\boldsymbol{d} = (k_c, k_s)$ to represent the colored degree of node $u$. Since $u$ have one type-$c$ edge connecting to a node at level $\ell + 2$, we haves
\vspace{-1mm}

\begin{equation}
    \label{eq:qcli}
    \begin{aligned}
    & q_{c, \ell+1}^i = \sum_{\boldsymbol{d} = (k_c, k_s)}  \frac{p_{\boldsymbol{d}} k_c}{\left\langle k_c\right\rangle}
    f_i(\boldsymbol{q}_{c,\ell}, \boldsymbol{q}_{s,\ell}, k_c - 1, k_s)
    \end{aligned}
\end{equation}
\begin{equation}
    \label{eq:qsli}
    \begin{aligned}
    & q_{s, \ell+1}^i = \sum_{\boldsymbol{d} = (k_c, k_s)}  \frac{p_{\boldsymbol{d}} k_s}{\left\langle k_s\right\rangle}
    f_i(\boldsymbol{q}_{c,\ell}, \boldsymbol{q}_{s,\ell}, k_c, k_s - 1)
    \end{aligned}
\end{equation}

For the probability that a type-$i$ root node at level $\infty$ to \textit{not} be infected, we can take expectation over the degree distribution of the node. Therefore

\begin{equation}
    \label{eq:qi}
    \begin{aligned}
    & q_{\infty}^i = 
    \lim_{\ell \to \infty}
    \sum_{\boldsymbol{d} = (k_c, k_s)} p_{\boldsymbol{d}} 
     f_i(\boldsymbol{q}_{c,\ell}, \boldsymbol{q}_{s,\ell}, k_c, k_s)
    \end{aligned}
\end{equation}

Recall $q_{\infty}^i$ represent the \textit{non-infection} probability of a random type-$i$ node at root. Conditioned on the event that the epidemic emerges, $1 - q_{\infty}^i$ is the probability that a uniform random type-$i$ node is eventually infected. The total fraction of infection given emergence is given by $\sum_{i=1}^M m_i (1 - q_{\infty}^i)$.

\section{Numerical Results}
We next present numerical simulations that validate our analytical results.
The contact networks are generated via the configuration model with Poisson degree distribution and 10,000 nodes for all the presented experiments. To generate the simulation plots, we took an average over 5,000 independent trials where, in each trial, a new contact network is generated. The threshold of epidemic emergence is 0.05 for all the experiments.

\subsection{Spread process as a function of mean degrees}
\label{exp30}
In this experiment, we assume there are two types of nodes in the population: mask wearers (type-1) and no-mask wearers (type-2).
The vector $\boldsymbol{m} = [m1, m2]$ denotes the percentages of the population for the two types of nodes. 
The inward efficiencies of the two types of masks is represented by vector $\boldsymbol{\epsilon}_{in} = [\epsilon_{in,1}, \epsilon_{in,2}]$, and outward efficiencies are given by $\boldsymbol{\epsilon}_{out} = [\epsilon_{out, 1}, \epsilon_{out, 2}]$.
For this experiment, we set $\boldsymbol{m} = [0.2, 0.8]$, $\boldsymbol{\epsilon}_{out} = [0.6, 0]$, $\boldsymbol{\epsilon}_{in} = [0.5, 0]$.
In this inward and outward efficiency set up, value 0 represents no mask protection at all.
$\alpha$ is roughly estimated as the average percentage of school-age kids in the households \cite{avg_num_kids, avg_familiy_size}. 
Consider that there might be adult school staffs, we conduct sensitivity analysis on the choice of $\alpha$ in Section \ref{exp38}. 

\begin{figure}[h]
    \centering
    \subfigure[Probability of Emergence]{
    \label{fig:md_prob}
    \includegraphics[width=0.2\textwidth]{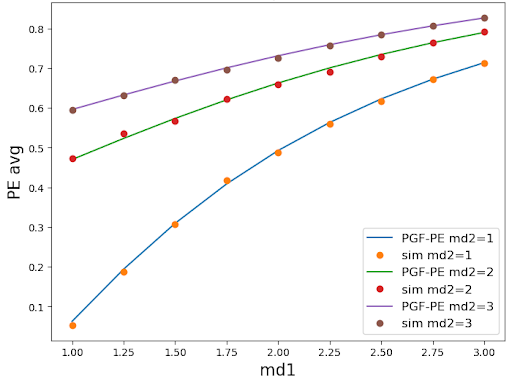}}
    \subfigure[Epidemic Size]{
    \label{fig:md_frac}
    \includegraphics[width=0.2\textwidth]{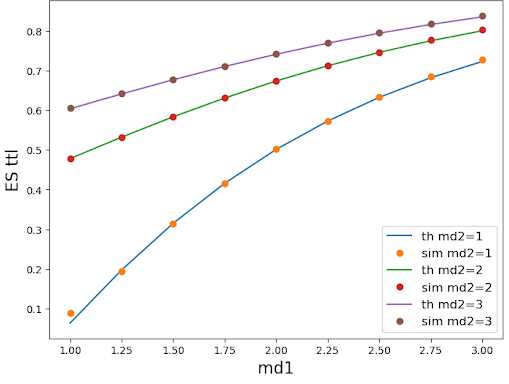}}
    \caption{\sl 
The average probability of the emergence (a), the total epidemic size given emergence (b) for mask and no mask on synthetic networks generated by the configuration model with Poisson degree distribution on varying mean degrees. $md1$ and $md2$ represent the mean degree of network $\mathbb{C}$ and network $\mathbb{S}$, respectively. $\boldsymbol{m} = [0.2, 0.8]$, $\boldsymbol{\epsilon}_{out} = [0.6, 0]$, $\boldsymbol{\epsilon}_{in} = [0.5, 0]$. 
}
    \label{fig:md_pe_es}
\end{figure}

Figure \ref{fig:md_pe_es} show the impact of mean degrees of each layer in the multi-layer network $\mathbb{H}$ on the probability of emergence of an epidemic, as well as the expected final epidemic fraction. 
$md_1$ and $md_2$ represent the mean degree of network $\mathbb{C}$ and network $\mathbb{S}$, respectively.
As we can see from figure \ref{fig:md_prob}, the simulation results match the analytical solutions near-perfect with only 10,000 nodes in the simulation. This validated our analytical solutions which is derived under asymptotic condition.
In addition, we can observe that for both PE and ES, as the $md_1$ increase their values increase. 
Also, higher $md_2$ values lead to higher both PE and ES.
Our results show that, mitigation strategies that keep the mean degrees of both networks low can effectively prevent the epidemic from happening, 
or reduce the final size of the epidemic if it already exists.


\subsection{Opening/closure of the school layer and mask-wearing}
\label{exp38}

In this experiment, we explore the impact of open/closure of the school layer with different population-level mask-wearing status.
In this experiment, we assume there are two types of nodes in the population: surgical mask wearers (type-1) and cloth mask wearers (type-2).
The vector $\boldsymbol{m} = [m1, m2]$ denotes the percentages of the population for the two types of nodes. 
The inward efficiencies of the two types of masks is represented by vector $\boldsymbol{\epsilon}_{in} = [\epsilon_{in,1}, \epsilon_{in,2}]$, and outward efficiencies are given by $\boldsymbol{\epsilon}_{out} = [\epsilon_{out, 1}, \epsilon_{out, 2}]$.
We set $md_1 = 6$ $md_2 = 8$ according to work done by \cite{noauthor_network-based_nodate, potter_estimating_2012}.
Vector $\boldsymbol{m} = [m_1, m_2]$ where $m_1 + m_2 = 1$ denotes the percentage of each mask type, and $\alpha$ denotes the percentage of population that are also in the school layer. 
We set $\boldsymbol{\epsilon}_{out} = [0.8, 0.5]$, $\boldsymbol{\epsilon}_{in} = [0.7, 0.5]$ based on \cite{tian_role_2022}.
The original transmissibility on each layer is set to: 
$T_c = 0.6$ and $T_s = 0.5$ based on \cite{cdcmmwr_coronavirus_2020}
We vary the values of $m1$ (and $m2$ will be of value $1 - m_1$) from $[0.1, 0.3, 0.5, 0.7, 0.9]$. 
In the meanwhile, we vary the values of $\alpha$ in the same way.
The bigger value of alpha represents higher level of openness of school layer.

\begin{figure}[h]
    \centering
    \subfigure[Probability of Emergence]{
    \label{fig:alpha_m_pe}
    \includegraphics[width=0.2\textwidth]{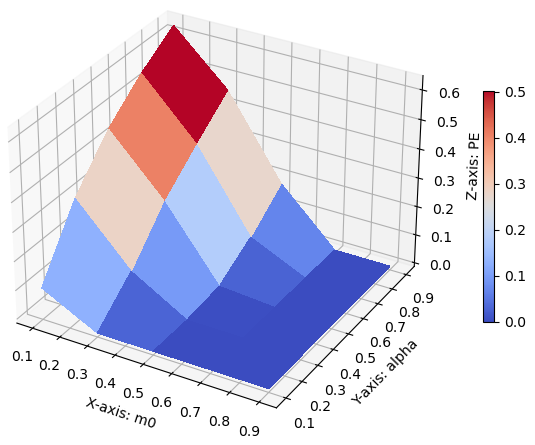}}
    \subfigure[Epidemic Size]{
    \label{fig:alpha_m_es}
    \includegraphics[width=0.2\textwidth]{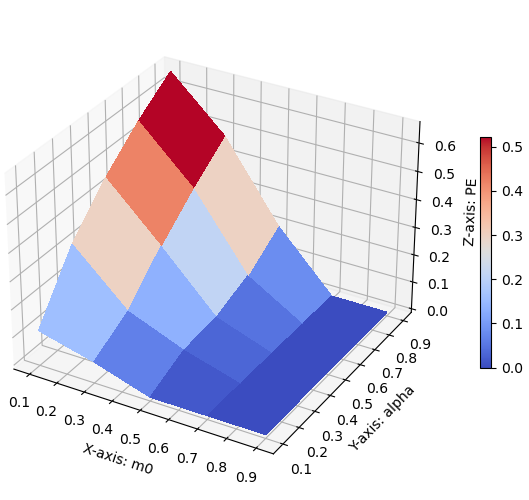}}
    \caption{\sl 
The average probability of the emergence (a), the total epidemic size given emergence (b) for cloth mask and surgical mask on synthetic networks generated by the configuration model with Poisson degree distribution on varying mean degrees. 
$md_1 = 6$, $md_2 = 8$ for network $\mathbb{C}$ and $\mathbb{S}$.
$\boldsymbol{\epsilon}_{out} = [0.8, 0.5]$, $\boldsymbol{\epsilon}_{in} = [0.7, 0.5]$.
$T_c = 0.6$ and $T_s = 0.5$.
}
    \label{fig:alpha_m}
\end{figure}

Figure \ref{fig:alpha_m} shows the results on probability of emergence, epidemic size as $\alpha$ and $m_0$ vary. 
We see that PE and ES are behaving the similar way as $\alpha$ and $m_0$ changes.
Interestingly, we can see that as $\alpha$ increases, with large $m_0$ values 0.7 and 0.9, PE and ES values remain zero.
With $m_0$ values smaller than 0.7, increase $\alpha$ indeed increases the risk of leading to an epidemic and expanding the epidemic size if the epidemic exists. 
This trend tells us that if in the real-life scenario, if \textit{proper} proportion of the population wear good-quality masks such as surgical masks, and the rest wear cloth masks, it's possible to suppress the viral transmission from happening at all even we open the school, or not make the current epidemic worse if we are already in one. 
Our model is able to obtain this important \textit{threshold} of proportion of masks for each type, providing important viral mitigation guidelines for policy decision making process.

\subsection{Trade-offs between inward and outward mask efficiency}
\label{exp50}
In this section, we validate the findings on the trade-off between source-control and self-protection on a single-layer network setting by \cite{tian_role_2022} on a multi-layer network setup.
Adopting the same experiment setup in \cite{tian_role_2022}, we assume there are three types of masks in the population: inward-good mask wearers, outward-good mask wearers and no mask wearers.
We also adopt the same parameter choice for the inward and outward efficiencies of the above three types of masks:
$\boldsymbol{\epsilon}_{in} = [0.7, 0.3, 0]$ and
$\boldsymbol{\epsilon}_{out} = [0.3, 0.7, 0]$. 
We keep the proportion of the no mask population as 0.1, thus we have $m_{\text{inward-good}} + m_{\text{outward-good}} = 0.9$.
And we vary the value of $m_{\text{outward-good}}$ from 0.1 to 0.7.

\begin{figure}[h]
    \centering
    \label{fig:alpha_m_inout}
    \includegraphics[width=0.25\textwidth]{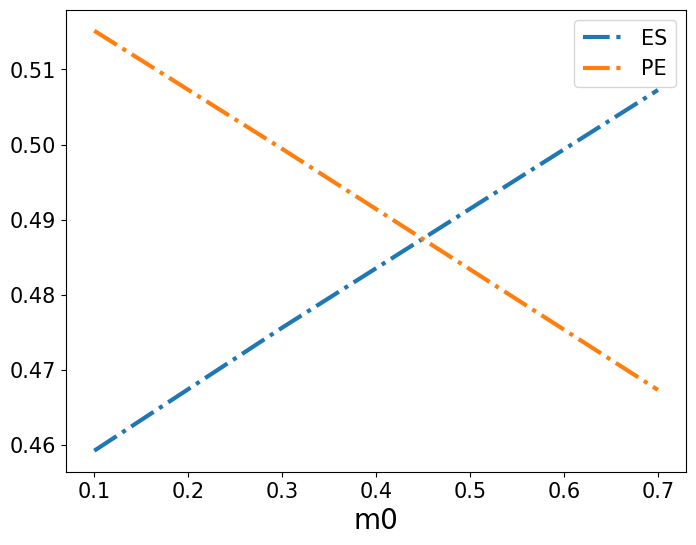}
    \caption{\sl 
The average probability of the emergence (yellow), the total epidemic size given emergence (blue) for inward-good mask, outward-good mask and no mask on synthetic networks generated by the configuration model with Poisson degree distribution on varying mean degrees. 
$md_1 = 6$, $md_2 = 8$ for network $\mathbb{C}$ and $\mathbb{S}$.
$\boldsymbol{\epsilon}_{in} = [0.7, 0.3, 0]$ and
$\boldsymbol{\epsilon}_{out} = [0.3, 0.7, 0]$. 
$T_c = 0.6$ and $T_s = 0.5$.
$m_0$ is the proportion of outward-good masks.}
\end{figure}

Unlike the results observed in figure \ref{fig:md_pe_es} and \ref{fig:alpha_m}, where PE and ES behave the same way as the variables vary, here PE and ES are exhibiting the opposite trend as $m_0$ increases. 
This trend is first described in \cite{tian_role_2022} by exploring the trade-off between source-control and self-protection over a single-layer contact network.
Here we conduct the same experiment on a multi-layer contact network and obtain the same finding.
As shown in figure \ref{fig:alpha_m_inout}, before the epidemic exists, outward-good masks (better efficiency at source-control) are more efficient to reduce the risk of an epidemic emergence. 
On the other hand, if the epidemic exists, inward-good masks are more useful to suppress the further expansion of the epidemic.
We see that spreading process should be considered as two phases rather than one either on single-layer, or on multi-layer networks.

\vspace{-1mm} 
\section{Conclusion}
In this paper, we have studied an agent-based model for viral spread on \textit{multi-layer} networks with \textit{population heterogeneity} called the \textit{multi-layer mask model}.
In the model, the heterogeneous viral transmission probability
between nodes in a node pair depends both on the node types and the type of link  that connect both nodes.
In particular, we perform a theoretical analysis of three important quantities: the probability of emergence, the epidemic threshold, and the expected epidemic size. 
We validated our analytical results by comparing them against agent-based simulations and they show a near-perfect match.
We found that it's safe to open the school layer with corresponding mask distribution over the population.
Last but not the least, we validate the conclusion by \cite{tian_role_2022} on the trade-off between source-control and self-protection over single-layer networks  on the multi-layer network setting. 
It demonstrates that when considering the mitigation strategies we need to treat the spreading process as two phases rather than one over the multi-layer networks as well.

\bibliography{references.bib}

\begin{thebibliography}{10}
\bibitem{noauthor_network-based_nodate}
``A network-based explanation of why most {COVID}-19 infection curves are
  linear {\textbar} {PNAS}.''
\end{thebibliography}
\bibliographystyle{ieeetr}
\vspace{12pt}
\end{document}